# Network evasion detection with Bi-LSTM model

Kehua chen[1], JingPing Jia[2]

[1] *North China Electric Power University, chenkehualve123@163.com, China*
[2] *North China Electric Power University, rocsovsky@163.com, China*

**Abstract.** Network evasion is a way to disguise data traffic by confusing network intrusion detection systems. Network evasion detection is designed to distinguish whether a network traffic from the link layer poses a threat to the network or not. At present, the traditional network evasion detection method does not extract the characteristics of network traffic and the detection accuracy is relatively low. In this paper, a novel network evasion detection framework has been proposed to detect eight atomic evasion behaviors which are based on deep recurrent neural network. Firstly, inter-packet and intra-packet features are extracted from network traces. Then a bidirectional long short-term memory (Bi-LSTM) neural network is trained to encode both the past and the future traits of the network traces. Finally, on the top of the Bi-LSTM network, a Softmax layer is used to classify the trace into the correct evasion class. The experimental results show that the average detection accuracy of the framework reaches 96.1%.

**Keywords:** Network evasion, Network Intrusion Detection System, Recurrent Neural Network.

## 1 Introduction

Evasions can be applied to normal traffic, as well as to attacks. However, advanced evasion is able to disguise attacks to avoid detection and blocking by the network security system. Evasions are considered successful as long as the delivery mechanism succeeds in gaining access to victim computers while the security device fails to detect or respond to the attack. However, there are not effective ways to detect network evasions. When faced with a large volume of network flows, existing methods have various deficiencies.

Cheng [1] described five common techniques that can evade the examination of an IPS. They are DOS, packet splitting, duplicate insertion, payload mutation and shell code mutation. Varghese and Antichi [2-3] proposed to cut the signature into splits and match them in the split signatures without reassembly, but their methods are still vulnerable to other evasion techniques. Zigang, XIONG [4] proposed a two-phased method for the detection of malware communication channels, but it's still limited to detect limited evasion types in a small TCP/IP network.

In this paper, we propose to tackle the evasion problem from the view of machine learning and present a network evasion detection framework based on deep recurrent neural network. First, we describe the features extracted from evasion flows. Then we take an overview of Long Short Term Memory Network (LSTM) [5] and Bidirectional Recurrent Neural Network (Bi-RNN). Both of them were proposed in 1997 by Mike Schuster and Kuldip K. Paliwal[6] which successfully had overcome the deficiency of long term dependencies.

## 2 Features

Normal network flows were extracted from an intranet network flows. Eight evasion techniques, which are ip_chaff, ip_frag, ip_opt, ip_ttl, ip_tos, tcp_chaff, tcp_opt and tcp_seg were applied to the normal network flows. Intra-packet and inter-packet features were extracted from trace files of each type of evasion technique to form samples of each evasion class. The choice of features aims to reflect the characteristics of each evasion technique. Table 1 shows the complete list of the features' names and their values. Intra-packet features which are extracted only from the current IP packet are iplen, ipflag, tcpchecksum, tcplen, tcpflag, tcpsyn, tcpseq, tcpmss, iproute, ipopt, iptos, tcpwscale and tcpack. Inter-packet features which are generated from two sequential packets are ipoffset, overlap, ipttldiff, tcpoverlap, are tcp timestamp. Table 2 shows the features extracted from an example TCP stream consisting of five packets.

**Table 1.** Features.

| **Dimension** | **Feature name** | **Feature value** |
|---|---|---|
| 0 | Iplen | The length of the IP packet |
| 1 | Ipoffsetoverlap | -1, if this is the first packet of the stream<br>Otherwise, fragment offset of the current packet - fragment offset of the previous |

[1] Kehua Chen: chenkehualve123@163.com

| | | |
|---|---|---|
| | | packet |
| 2 | ipflag | The value of the flags field in the IP header |
| 3 | ipttldiff | 1, if this is the first packet of the stream<br>Otherwise, the value of the ttl field in the IP header of current packet - the value of the ttl field in the IP header of previous Packet |
| 4 | Tcpchecksum | -1, if current IP packet does not contain TCP segment<br>-2, if the TCP header in the current IP packet does not contain checksum field<br>0, if the checksum field of the TCP header in the current IP packet shows Wrong<br>Otherwise,1, the checksum field of the TCP header in the current IP packet shows Correct |
| 5 | tcplen | -1, if the current IP packet does not contain TCP segment<br>-2, if the length field of the TCP header in the current IP packet is invalid<br>Otherwise, the value of the length field of the TCP header |
| 6 | Tcpoverlap | 1, if the current IP packet does not contain TCP segment and previous packet does not have effective TCP sequence number or segment length<br>2, if the current IP packet does not contain TCP segment but previous packet has effective TCP sequence number and segment length<br>3, if the current IP packet contains TCP segment but previous packet does not have effective TCP sequence number or segment length<br>-6, if the current IP packet contains TCP segment but does not have effective TCP sequence number or segment length<br>-1, if the TCP segment in the current IP packet ends before the TCP segment in the previous IP packet starts<br>-2, if the TCP segment in the current IP packet starts before and ends before or the same with the TCP segment in the previous IP packet<br>-3, if the TCP segment in the current IP packet starts before or the same with, and ends after the TCP segment in the previous IP packet<br>-4, if the TCP segment in the current IP packet starts after and ends before or the same with the TCP segment in the previous IP packet<br>-5, if the TCP segment in the current IP packet starts after or the same with, and ends after the TCP segment in the previous IP packet<br>Otherwise,0, the TCP segment in the current IP packet starts after the TCP segment in the previous IP packet ends. |
| 7 | tcpflag | -1, if the current IP packet does not contain TCP segment<br>-2, if the TCP segment in the current IP packet does not have a valid flag field<br>Otherwise, the value of the flag field of the TCP segment in the current IP packet |
| 8 | Tcptimestamp | -1, if the current IP packet does not contain TCP segment or it is the first packet in the stream<br>-2,if the TCP segment in the current IP packet does not have a valid timestamp<br>Otherwise, timestamp value of the TCP segment in the current IP packet minus the timestamp value of the previous IP packet |
| 9 | Tcpsyn | -1, if the current IP packet does not contain TCP Segment<br>1, if the TCP segment in the current IP packet has SYN flag<br>Otherwise, 0 |
| 10 | Tcpseq | -1, , if the current IP packet does not contain TCP Segment<br>-2, if the TCP segment in the current IP packet does not have a valid sequence number<br>-3, if the TCP header in the previous packet from the other side does not specify window size.<br>0, if the sequence number of the TCP segment is within the TCP window specified by the TCP header in the previous IP packet from the other side.<br>Otherwise, 1 |
| 11 | Tcpmss | -1, if the current IP packet does not contain TCP segment.<br>0, if the TCP header in the current IP packet does not have mss option.<br>Otherwise,1 |
| 12 | Iproute | -1, if the current IP packet does not have the route Option<br>0, if the current IP packet has the strict source and record route(SSRR) option<br>1, if the current IP packet has the loose source and record route(LSRR) option<br>Otherwise, 0 |
| 13 | Ipopt | 0, if the current IP packet's header has a valid option field<br>Otherwise,1, the option field in the current IP packet's header has an undefined value. |
| 14 | Iptos | -1, if the header of the current IP packet does not set TOS field<br>Otherwise, value of the TOS field in the header of the current IP packet |
| 15 | Tcpwscale | -1, if the current IP packet does not contain TCP segment.<br>-2, if the TCP header in the current IP packet does not have window scale option.<br>Otherwise,1 |
| 16 | Tcpack | -1, if the current IP packet does not contain TCP segment.<br>0, if the TCP segment in the current IP packet has ACK flag<br>Otherwise,1 |

**Table 2.**Example features from a TCP stream consisting of six packets.

| Dimension | Packet No | | | | | |
|---|---|---|---|---|---|---|
| | 0 | 1 | 2 | 3 | 4 | 5 |
| 0 | 48 | 40 | 364 | 364 | 40 | 40 |
| 1 | -1 | 0 | 0 | 0 | 0 | 0 |
| 2 | 2 | 2 | 2 | 2 | 2 | 2 |
| 3 | 1 | 0 | -121 | 121 | 0 | 0 |
| 4 | -2 | -2 | -2 | -2 | -2 | -2 |
| 5 | 0 | 0 | 324 | 324 | 0 | 0 |
| 6 | 3 | 0 | 0 | -4 | 0 | 0 |
| 7 | 2 | 16 | 24 | 24 | 17 | 16 |
| 8 | -1 | -2 | -2 | -2 | -2 | -2 |
| 9 | 1 | -2 | -2 | -2 | -2 | -2 |
| 10 | -3 | 0 | 0 | 0 | 0 | 0 |
| 11 | 1 | 0 | 0 | 0 | 0 | 0 |
| 12 | -1 | -1 | -1 | -1 | -1 | -1 |
| 13 | 0 | 0 | 0 | 0 | 0 | 0 |
| 14 | 0 | 0 | 0 | 0 | 0 | 0 |
| 15 | -2 | -2 | -2 | -2 | -2 | -2 |
| 16 | 0 | 1 | 1 | 1 | 1 | 1 |

## 3 Evasion Detection Model

In our evasion detection model, a deep recurrent neural network based on LSTM is built to encode the feature sequences extracted from network flows. As a variant of RNN, LSTM just replaces RNN cells with LSTM cells. A bidirectional recurrent neural network (bi-RNN) is an extension of RNN that learns the input information twice from left to right and from right to left. Bidirectional LSTM (Bi-LSTM) is a bi-RNN counterpart based on LSTM.

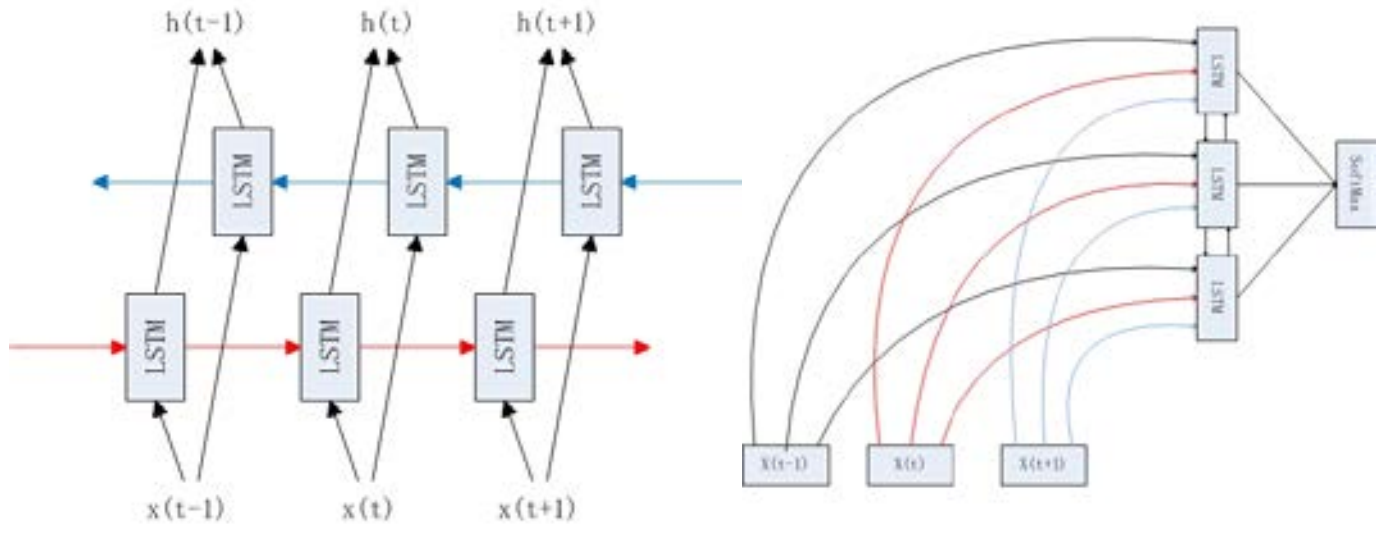

(a) (b)

**Figure 1**. The Bidirectional LSTM's structure and the Bi-LSTM architecture

As described previously, an LSTM cell only has a forward conduction and each unit is influenced only by the previous unit. However, in network evasion, the future part of information is equally important with the past one. The structure of a Bi-LSTM shows in Figure 1-(a). In that way, Bi-LSTMs can effectively make use of past states and future states in a specific period.

Specifically, we adopt Bi-LSTM combining the Softmax [7] for evasion detection as illustrated in Figure 1-(b). The input is a feature sequence of time-series from a series of evasion, $(x_{(1)}, x_{(2)}, \cdots, x_{(t)})$, the input is divided into a fixed structure and accordingly fed to the LSTM unit one by one. When finishing feeding one batch size network flow, the whole network model begins the backpropagation to complete training.

## 4 Experiments

### 4.1 Dataset

Experiments were carried out based on the databases crawled on the network. We transformed crawled data and then saved the features into numpy files separately. Each file contains a feature sequence with its length ranging from 3 to 7, holding 16 dimensions. We randomly split them into a training set with 183411 numpy files and a testing set with 45853 numpy files for subsequent works.

### 4.2 Experimental setup

To update the network parameters fast, we set the train batch size to 50, the frame size to 5, and the number of hidden layer cells to 128 with Adam optimizer initially. The experimental results show that the Adam optimizer can ensure stable convergence and average accuracy of 97.01%. Since the dataset is large enough, there is no need to apply dropout. The confusion matrix of the network evasion detection is shown in Table 3 and the ROC (Receiver Operating Characteristic) curve is shown in Figure 2.

**Table 3.**Confusion Matrix for 8 evasion classes in Bi-LSTM approaches

| Confusion Matrix for network evasion detection in 8 classes | | | | | | | | | |
|---|---|---|---|---|---|---|---|---|---|
| | IP_ Opt | IP_ Frag | TCP_ chaff | IP_ Tos | TCP_ seg | IP_ ttl | IP_ chaff | TCP_ opt | Accura Rate(% |
| IP_opt | 4177 | 0 | 0 | 0 | 0 | 0 | 0 | 161 | 96.29 |
| IP_frag | 0 | 12249 | 17 | 0 | 1 | 47 | 0 | 58 | 99.01 |
| TCP_chaff | 0 | 0 | 4995 | 0 | 68 | 39 | 0 | 184 | 94.49 |
| IP_tos | 0 | 0 | 0 | 3867 | 0 | 0 | 0 | 149 | 96.29 |
| TCP_seg | 0 | 93 | 15 | 0 | 3064 | 66 | 0 | 98 | 91.85 |
| IP_ttl | 0 | 136 | 16 | 0 | 45 | 3303 | 0 | 150 | 90.49 |
| IP_chaff | 0 | 0 | 1 | 0 | 0 | 0 | 4095 | 2 | 99.93 |
| TCP_opt | 0 | 0 | 0 | 0 | 0 | 0 | 0 | 7984 | 100 |

**Figure 2**. ROC curve for Network evasion detection

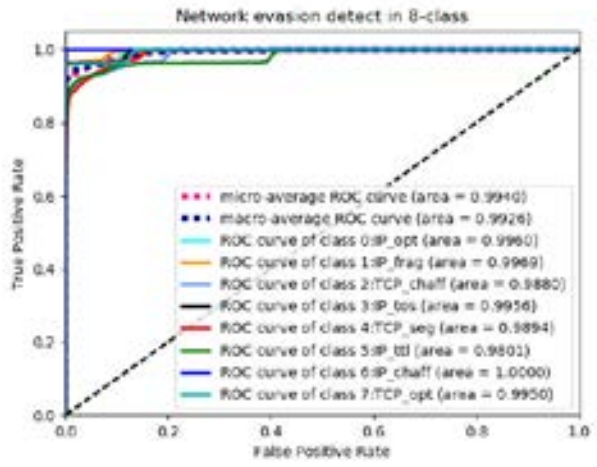

### 4.3 Learning Rate and Batch size

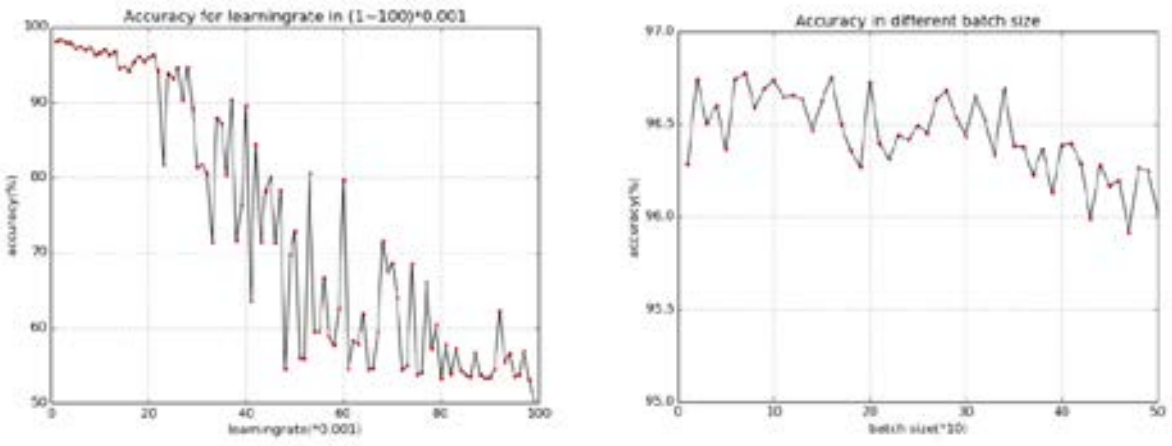

(a) (b)

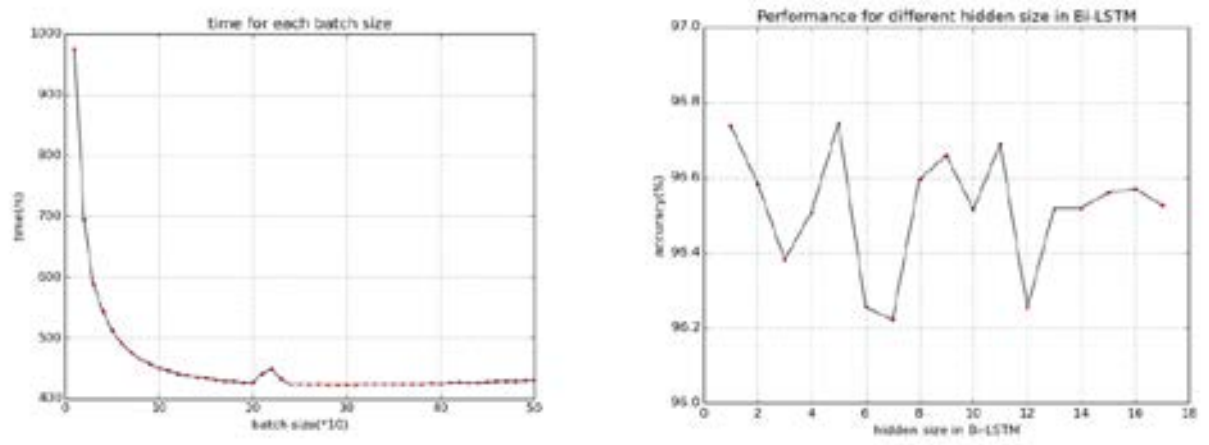

(c) (d)

**Figure 3-(a),(b),(c),(d)** Performance in different learning rate, batch size, in different hidden size in Bi-LSTM, the running time in different batch size.

The learning rate determines the speed which the parameter moves to the optimal value. Excessive learning rates probably lead to consequences missing the optimal parameters, while with lower learning rates, optimization efficiency is likely to be unsatisfactory and algorithm takes a long time (as shown in Table 2 and Figure 3) to converge.

A fixed two-layer network was chosen, and different numbers of hidden cells were tested, but without observable differences. However, it is obvious that the higher the batch size in the experiment is, the shorter the required time it needs. As shown in Figure 3-(d), using different numbers of hidden layers does not cause significant differences.

### 4.4 Optimization methods and Regularization methods

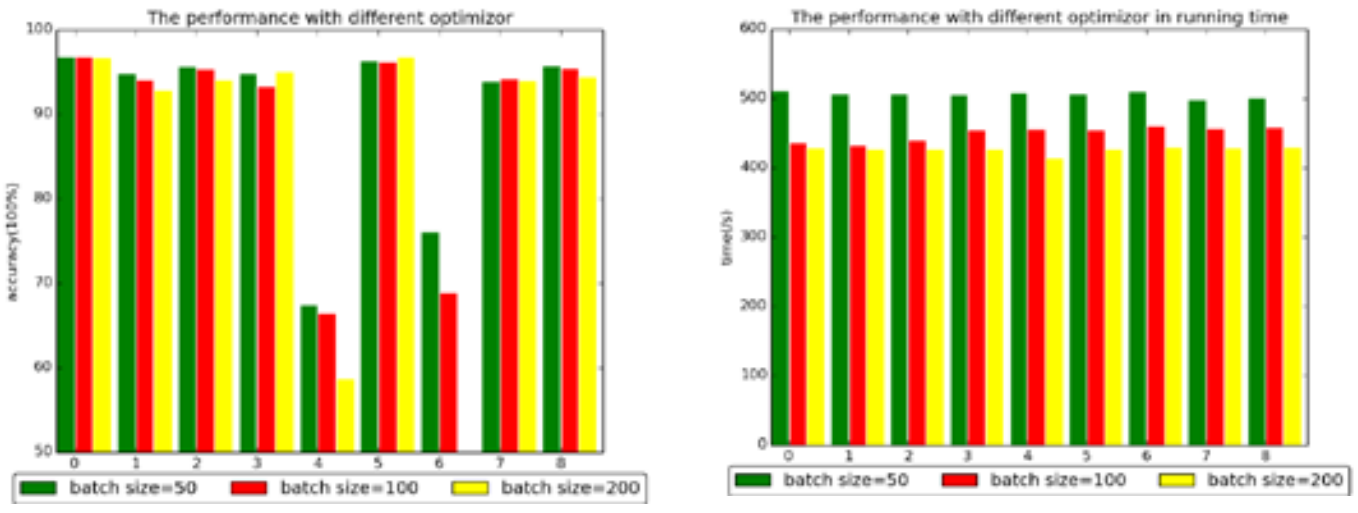

(a) (b)

**Figure 4-(a)** Performance in different batch size with different optimizer in Bi-LSTM, **Figure 4-(b)** Running time in different batch size with different optimizer in Bi-LSTM(0:Adam,1:GradientDescent,2:Adagrad,3:Momentum,4:Adadelta,5:RMS,6:Ftrl,7:ProximalGradientDescent, 8:ProximalAdadelta)

**Table 4.** Accuracy with different other hyper-parameters.2 layers, hidden layer size 50, iteration = 6.6×105.

| method | Lr | Dropout | L3 | L4 | L5 | L6 | L7 |
|---|---|---|---|---|---|---|---|
| GD | $10^{-3}$ | 0 | 85.11 | 92.64 | 94.46 | 94.17 | 94.67 |
| | $10^{-3}$ | 0.01 | 84.46 | 92.49 | 93.36 | 91.15 | 93.53 |
| | $5\times10^{-3}$ | 0.01 | 83.79 | 92.21 | 92.87 | 91.83 | 93.01 |
| Momentum | $10^{-3}$ | 0 | 84.04 | 92.81 | 93.92 | 93.98 | 94.38 |
| | $10^{-3}$ | 0.01 | 83.72 | 92.63 | 93.43 | 92.83 | 93.22 |
| | $5\times10^{-3}$ | 0.01 | 81.71 | 91.90 | 93.03 | 92.82 | 92.05 |
| Adagrad | $10^{-3}$ | 0 | 86.09 | 94.41 | 95.63 | 96.24 | 96.08 |
| | $10^{-3}$ | 0.01 | 84.63 | 93.53 | 94.92 | 94.74 | 95.02 |
| | $5\times10^{-3}$ | 0.01 | 85.75 | 95.08 | 95.92 | 96.17 | 96.51 |
| Adadelta | $10^{-3}$ | 0 | 63.27 | 67.62 | 68.77 | 72.13 | 78.46 |
| | $10^{-3}$ | 0.01 | 58.27 | 64.81 | 65.81 | 69.14 | 69.29 |
| | $5\times10^{-3}$ | 0.01 | 73.35 | 80.98 | 87.09 | 84.13 | 84.38 |
| RMS | $10^{-3}$ | 0 | 86.22 | 95.77 | **96.84** | 96.57 | 96.34 |
| | $10^{-3}$ | 0.01 | 85.97 | 95.21 | 96.37 | 97.02 | 96.78 |
| | $5\times10^{-3}$ | 0.01 | 85.45 | 94.87 | 96.33 | 95.92 | 93.23 |
| Ftrl | $10^{-3}$ | 0 | 57.33 | 61.44 | 76.37 | 78.36 | 82.66 |
| | $10^{-3}$ | 0.01 | 60.81 | 62.07 | 78.07 | 79.44 | 77.64 |
| | $5\times10^{-3}$ | 0.01 | 85.39 | 94.32 | 95.53 | 95.65 | 96.09 |
| Proximal GD | $10^{-3}$ | 0 | 84.73 | 93.48 | 95.30 | 93.84 | 94.67 |
| | $10^{-3}$ | 0.01 | 84.02 | 92.52 | 94.24 | 93.10 | 92.75 |
| | $5\times10^{-3}$ | 0.01 | 84.10 | 91.56 | 93.78 | 91.36 | 91.59 |
| Proximal Adagrad | $10^{-3}$ | 0 | 86.21 | 94.48 | 95.72 | 95.58 | 95.53 |
| | $10^{-3}$ | 0.01 | 84.63 | 93.82 | 95.12 | 94.85 | 94.69 |
| | $5\times10^{-3}$ | 0.01 | 86.04 | 94.94 | 96.16 | 96.07 | 96.66 |
| Adam | $10^{-3}$ | 0 | 86.09 | 95.24 | 96.73 | 96.93 | 96.79 |
| | $10^{-3}$ | 0.01 | 85.61 | 95.59 | 96.94 | 96.73 | 96.98 |
| | $5\times10^{-3}$ | 0.01 | 85.85 | 93.02 | 95.77 | 96.05 | 95.94 |
| | $10^{-2}$ | 0.01 | 84.15 | 92.55 | 94.95 | 94.82 | 95.89 |
| | $10^{-3}$ | 0.05 | 86.02 | 95.21 | 96.46 | 95.91 | 96.72 |

Seemingly the Adam and RMS out-perform other optimizers whatever lr(learning rate) and dropout are. In all training results, the classification accuracy is the lowest when the frame size is 3, which is caused by insufficient information. There is a slightly degradation when frame size comes to 7, which is because the original files were scarce and lots of data were dropped out. We also tried both dropout and standard L2 regularization in the initial experiments. It is trustable that dropout can beat the over-fitting when a larger dataset is accessible in the future.

## 5 Conclusions

In this paper, we addressed the network evasion detection problem from the view of sequence classification task. Specifically, we proposed to use Bi-LSTM network to encode both past and future features of the produced evasion network flows and classify new network traces to correct evasion type. Conclusions are as follows: 1) Bi-LSTM significantly outperforms unidirectional LSTM in network evasion detection. 2) Model with one or two layers achieves superior performance. 3) The performance varies on different tricks such as learning rate, optimization and regularization method. Further study may consider the detection of combination of two or more atomic evasions.